\documentclass[conference,a4paper]{IEEEtran}
\usepackage[left=1.43cm,right=1.43cm,top=1.82cm,bottom=4.3cm]{geometry}

\usepackage[algo2e]{algorithm2e} 
\usepackage{amsmath,amssymb,amsmath,amsfonts} 
\usepackage{bm} 
\usepackage{graphicx}
\usepackage{cite} 
\usepackage{epstopdf} 
\usepackage{gensymb}
\usepackage{color} 
\usepackage{lipsum} 
\usepackage{float} 
\usepackage{epstopdf} 
\usepackage[mathcal]{eucal}
\usepackage{subfiles} 
\usepackage[T1]{fontenc}
\usepackage{siunitx}
\usepackage[unicode=true,bookmarks=false,bookmarksnumbered=false,bookmarksopen=false, breaklinks=false,pdfborder={0 0 0},backref=false,colorlinks=false]{hyperref}
\usepackage{tabulary} \usepackage{epsfig,graphics,graphicx,subfig,amssymb,amstext,amsmath,algorithm,algorithmic,wrapfig,multirow}\usepackage{array}
\usepackage{adjustbox} 
\usepackage[dvipsnames]{xcolor} 
\usepackage{placeins} 
\usepackage{textcomp} 
\usepackage[font=small, skip=4pt]{caption} 
\usepackage{flushend} 
\usepackage{color, colortbl}
\usepackage{tabularx}
\usepackage{enumerate} 
\usepackage{tikz} 
\usetikzlibrary{shapes,arrows} 
\usepackage[utf8]{inputenc} 
\usepackage{mathtools} 
\usepackage{xcolor} 
\usepackage{tikz} 
\usetikzlibrary{chains,arrows,calc,positioning}
\usepackage{amssymb}
\usepackage{pifont}
\usepackage{soul} 
\usepackage{breqn} 
\usepackage{booktabs} 
\usepackage{enumitem} 
\usepackage{tabulary}
\usepackage[normalem]{ulem} 
\usepackage{dblfloatfix} 
\usepackage{makecell} 
\usepackage{lipsum}
\usepackage{amsthm} 
\usepackage{comment}

\usepackage{textcomp} 
\usepackage{xcolor} 
\usepackage{pifont} 
\usepackage{multicol} 
\usepackage{upgreek} 
\usepackage[english]{babel}

\newtheoremstyle{mystyle}
  {}
  {}
  {\itshape}
  {}
  {\bfseries}
  {.}
  { }
  {}
\theoremstyle{mystyle}

\newlength \figwidth
\definecolor{bittersweet}{rgb}{1.0, 0.44, 0.37}
\definecolor{glaucous}{rgb}{0.38, 0.51, 0.71}
\definecolor{gainsboro}{rgb}{0.86, 0.86, 0.86}
\definecolor{babyblueeyes}{rgb}{0.63, 0.79, 0.95}
\definecolor{silver}{rgb}{0.75, 0.75, 0.75}
\definecolor{neoncarrot}{rgb}{1.0, 0.64, 0.26}
\definecolor{Gray}{gray}{0.9}
\definecolor{LightCyan}{rgb}{0.88,1,1}
\definecolor{BackgroundLightBlue}{rgb}{0.97,0.97,1}
\definecolor{BackgroundGray}{gray}{0.98}

\makeatletter

 \let\oldforeign@language\foreign@language
 \DeclareRobustCommand{\foreign@language}[1]{%
   \lowercase{\oldforeign@language{#1}}}

\def\nb0{{\mathbf{0}}}
\def\nb1{{\mathbf{1}}}

\makeatother

\IEEEoverridecommandlockouts

\begin{document}

\bstctlcite{IEEEexample:BSTcontrol}

\title{Autoencoder-Based CSI Compression for Beyond Wi-Fi 8 Coordinated Beamforming}

\author{

   \IEEEauthorblockN{
        Ibrahim Abou Shehada\IEEEauthorrefmark{1},
        Boris Bellalta\IEEEauthorrefmark{1},
        Giovanni Geraci\IEEEauthorrefmark{1}\IEEEauthorrefmark{2},
        Lorenzo Galati Giordano\IEEEauthorrefmark{3}
    }
    \\    \vspace{-0.3cm}

    \IEEEauthorblockA{
        \IEEEauthorrefmark{1}%
        Department of Engineering, Universitat Pompeu Fabra, Barcelona, Spain \\
        \IEEEauthorrefmark{2}%
        Nokia Standards, Spain \\
        \IEEEauthorrefmark{3}%
        Radio Systems Research, Nokia Bell Labs, Stuttgart, Germany \\
        Email: \{ibrahim.aboushehada@upf.edu\}
    } 
    \vspace{-0.5cm}

\thanks{This work was supported in part by grants PID2024-155470NB-I00 (MCIU/AEI/FEDER,UE), PCI2023-145958-2 (MLDR ANR-23-CHR4-0005), PID2021-123999OB-I00, PID2021-123995NB-I00, PID2024-156488OB-I00, CEX2021-001195-M, CNS2023-145384 and AGAUR ICREA Academia 00077.}
    
} 

\maketitle

\begin{abstract}
Coordinated beamforming (Co-BF) is a key multi-access-point coordination (MAPC) technique for dense Wi-Fi deployments, but its performance can be hindered by the large channel state information (CSI) feedback required through channel sounding across overlapping basic service sets (OBSS). This work proposes an autoencoder (AE)-based CSI compression mechanism integrated into a standards-aligned IEEE 802.11bn MAC design. Using an event-driven simulator with realistic channels generated through Sionna RT, we evaluate the tradeoff between AE reconstruction accuracy and feedback size by measuring their impact on channel sounding overhead and data latency. Our results show that AE-based compression reduces channel sounding overhead by more than 50\% compared to IEEE 802.11 CSI compression, with a compression ratio of 1/4 providing the best accuracy/feedback-size tradeoff for lowest data latency. Compared to legacy transmissions without MAPC, IEEE 802.11 CSI compression limits Co-BF due to high channel sounding overhead, causing it to underperform the legacy in some situations. However, AE-based CSI compression enables better Co-BF performance with substantial gains in throughput and data latency compared to legacy, demonstrating its promise as an enabler of efficient MAPC operation in future Wi-Fi systems.
\end{abstract}
\section{Introduction}
\label{sec:intro}

Wi-Fi usage continues to grow rapidly, driven by applications with high data rate, low latency, and high reliability requirements such as augmented reality/virtual reality (AR/VR), 4K/8K video streaming, and cloud gaming~\cite{gio2025wifitwentyfiveyearscounting}. However, contention-based channel access using carrier sense multiple access with collision avoidance (CSMA/CA) in IEEE 802.11 limits the ability to serve many highly-demanding users simultaneously in dense deployments \cite{Zhang2024}.

Multi-access-point coordination (MAPC) is a key feature anticipated for next-generation Wi-Fi (Wi-Fi 8 / ultra-high reliability (UHR) / IEEE 802.11bn)~\cite{ieee802.11bn}. With MAPC, when an access point (AP) gains channel access via CSMA/CA, it can share time, frequency, or spatial resources with another AP during the same transmission opportunity (TXOP), improving channel access in overlapping basic service sets (OBSS)~\cite{Giordano2024, Verma2024}. Coordinated beamforming (Co-BF) is a promising MAPC technique due to its interference-mitigation capabilities, but it requires each AP to acquire channel state information (CSI) from both its own stations (STAs) and OBSS STAs, resulting in large CSI feedback overhead that can negatively impact performance.

Prior Co-BF studies remain limited. The work in~\cite{Ezri2023} demonstrates zero multiuser interference (MUI) but only for one BSS, leaving inter-BSS interference unsuppressed for the other BSS. The analysis in~\cite{Strobel2024-2} focuses only on throughput, and~\cite{Jamshid2025} explores deep reinforcement learning (DRL) for Co-BF but uses simplified channel modeling with free-space path loss (FSPL) and shadowing. Most prior work assumes full-buffer traffic and relies on overly simplified channels, which can lead to exaggerated performance in multiuser (MU) multiple-input multiple-output (MIMO) settings, plus ignoring latency evaluation. Furthermore, the impact of channel sounding overhead on Co-BF performance remains unexplored.

In this work, we propose the use of an autoencoder (AE) for CSI compression to mitigate the negative impact of the large channel sounding overhead in Co-BF. We simulate Co-BF and legacy transmissions using a custom simulator built on SimPy~\cite{simpy} for event-driven simulation and Sionna Ray-Tracer (RT)~\cite{hoydis2023sionna} for realistic channel modeling, performing end-to-end (E2E) performance evaluations in terms of data latency and throughput. AE-based CSI compression has been studied in several works, such as CsiNet~\cite{csinet2019} and CRNet~\cite{crnet2020} for cellular communication and LB-SciFi~\cite{Gorinsky_lb_scifi2020} and EFNet~\cite{efnet2024} for Wi-Fi networks. To our knowledge, however, this is the first work that evaluates AE performance in terms of latency and investigates the optimal compression ratio based on E2E performance. First, we analyze the tradeoff between AE reconstruction accuracy and feedback size across different compression ratios and show that AE-based CSI compression reduces Co-BF channel sounding overhead by more than 50\% compared to the IEEE~802.11 method. In addition, an optimal tradeoff is achieved at a compression ratio of 1/4, which yields the lowest data latency. Furthermore, compared to legacy transmission without MAPC, Co-BF yields higher data latency in some scenarios using IEEE 802.11 CSI compression due to large sounding overhead, but the use of AE-based CSI compression allows Co-BF to outperform the legacy in all scenarios and offers substantial performance gains that can exceed 30\% in both data latency and throughput.
\begin{figure*}[t]
    \centering
    \includegraphics[width=\textwidth]{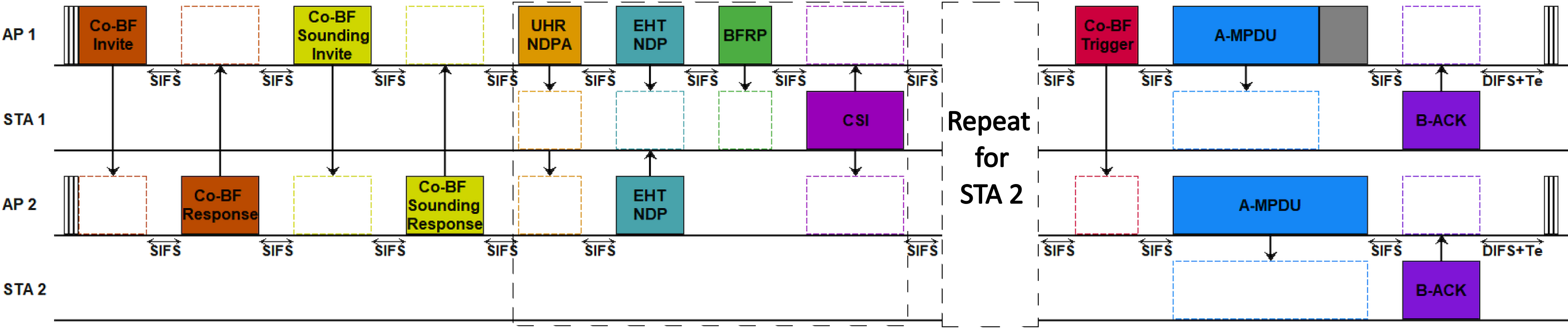}
    \caption{TXOP signaling structure example with 2 STAs for Co-BF with IEEE 802.11bn joint NDP channel sounding procedure.}
    \label{fig:mac}
\end{figure*}

\section{System Model}
\label{sec:sys}

In this section, details of the system model are discussed. In particular, we present the MAC layer model, channel model, formulation of the Co-BF precoder, and traffic model.

\subsection{MAC Layer Model}
\label{subsec:mac}



Each AP contends for the channel using the distributed coordination function (DCF). Once an AP gains channel access, it sends a Co-BF invite frame to the neighboring AP, which replies with a Co-BF response after a short inter-frame spacing (SIFS) interval (see Fig.~\ref{fig:mac}). The Co-BF invite and response frames follow the buffer state report poll (BSRP) non-trigger-based (NTB) and multi-STA BlockACK (BA) formats, respectively, as recommended by the IEEE802.11bn task group (TGbn)~\cite{ieee802.11bn}. We will refer to the initiating AP as the coordinating AP, and the other as the coordinated AP. Scheduling decisions are performed by the coordinating AP based on the oldest packets across both BSSs and are embedded in the Co-BF trigger frame, with a maximum of 2 scheduled STAs per BSS in the same TXOP. Both APs then transmit aggregated MAC protocol data unit (A-MPDU) frames to their scheduled STAs using Co-BF to suppress intra- and inter-BSS interference. If the coordinated AP has an empty buffer, it declines the session, and the coordinating AP starts a legacy A-MPDU transmission. Outdated CSI of any scheduled STA(s) triggers channel sounding prior to A-MPDU transmission, and we consider CSI to be outdated after 25~ms from the last update. This CSI age limit was specified based on the work in~\cite{Endovitskiy2024} and our own simulations to ensure that STAs can always receive data from their APs even if at the lowest modulation and coding scheme (MCS). \par

Co-BF sounding uses the IEEE 802.11bn joint null data packet (NDP) procedure~\cite{ieee802.11bn}. After the Co-BF response, the coordinating AP sends a sounding invite followed by the coordinated AP’s response. The coordinating AP then sends a UHR NDP announcement (NDPA), after which both APs simultaneously transmit Extremely-High Throughput (EHT) NDP frames. A beamforming report poll (BFRP) frame triggers uplink CSI feedback from the STAs. The coordinated AP repeats the NDPA/NDP/BFRP sequence for its own STA(s). If enough time remains in the TXOP, a Co-BF trigger initiates A-MPDU transmissions. Legacy transmissions without MAPC use the standard request-to-send/clear-to-send (RTS/CTS) procedure, and the MAC layer parameters are summarized in Table~\ref{tab:mac}. \par

\begin{table}[b]
	\centering
	\caption{MAC layer parameters}
	\label{tab:mac}
	\begin{tabular}{|m{.6\columnwidth}|c|}
    		\hline
    		\centering\textbf{Parameter} & \textbf{Value} \\
    		\hline
    		Maximum TXOP duration ($T_s$) [ms]  & 5.484 \\
            Minimum contention window size ($\text{CW}_{\text{min}}$) & 16 \\
            Maximum contention window size ($\text{CW}_{\text{max}}$) & 1024 \\
    		Payload length ($L_{D}$) [B] & 1500 \\
    		MAC header length [b] & 240 \\
    		MPDU delimiter [b] & 32 \\
    		Service field length [b] & 16 \\
    		Packet tail [b] & 18 \\
    		EHT/UHR Preamble duration [$\mu$s] & 88.8 \\
            Legacy preamble duration [$\mu$s] & 20 \\
            Legacy rate [Mbps] & 6 \\
            Short inter-frame spacing ($T_{\text{SIFS}}$) [$\mu$s] & 16 \\
            Distributed inter-frame spacing ($T_{\text{DIFS}}$) [$\mu$s] & 34 \\    
            Empty slot duration ($T_e$) [$\mu$s] & 9 \\
    		\hline
    	\end{tabular}
\end{table}

\subsection{Channel Model}
\label{subsec:phy}

Let $\mathbf{h}^k_{i,j} \in \mathbb{C}^{N_t}$ denote the channel between STA $i$ and AP $j$ in subcarrier $k$. The channels are generated using Sionna RT, and imperfect CSI is modeled by adding least-squares (LS) estimation noise to $\mathbf{h}^k_{i,j}$. During joint NDP sounding, STA $i$ receives long training field (LTF) pilot symbols from both APs as follows
\begin{equation}
y^k_{\text{LTF}_{i}} = \mathbf{h}^k_{i,1}\mathbf{x}^k_{\text{LTF}_1} + \mathbf{h}^k_{i,2}\mathbf{x}^k_{\text{LTF}_2} + n,
\end{equation}
where $n\sim\mathcal{CN}(0,N_0)$ is a complex white Gaussian noise with $N_0$ being the thermal noise power and the pilot symbols $\mathbf{x}^k_{\text{LTF}_1}$ from AP 1 and $\mathbf{x}^k_{\text{LTF}_2}$ from AP 2 are assumed to be orthogonal. To estimate $\mathbf{h}^k_{i,j}$, $y^k_{\text{LTF}_{i}}$ is multiplied by the pseudo-inverse of $\mathbf{x}^k_{\text{LTF}_j}$, which is orthogonal to $\mathbf{x}^k_{\text{LTF}_i}$, yielding
\begin{equation}
\mathbf{\hat{h}}^k_{i,j}
= y^k_{\text{LTF}_{i}}(\mathbf{x}^k_{\text{LTF}_j})^\dagger
= \mathbf{h}^k_{i,j} + \mathbf{n}_{\text{LTF}},
\end{equation}
where $\mathbf{n}_{\text{LTF}} \sim \mathcal{CN}(0,\sigma^2_{\text{LTF}}) \in \mathbb{C}^{N_t}$ is the LS estimation noise approximated as a complex Gaussian random vector with a variance of
\begin{equation}
\sigma^2_{\text{LTF}} = \frac{N_0 \cdot \text{NF}}{N_{\text{LTF}} \cdot E[\mathbf{x}_{\text{LTF}}^2]},
\end{equation}
where NF is the noise figure and $N_{\text{LTF}}$ is the number of pilot LTF symbols in the EHT NDP frame.

\subsection{Co-BF Precoder}
\label{subsec:cbf}

Co-BF is implemented using cell-edge-aware zero forcing (CEA-ZF)~\cite{Giovanni2017} to null interference between all scheduled STAs in both BSSs, assuming single-antenna receivers. For each AP–STA pair and subcarrier, we apply singular value decomposition (SVD) to the estimated channel as follows:
\begin{equation}
\mathbf{\hat{h}}^k_{i,j} = u^k_{i,j} \mathbf{s}^k_{i,j} \mathbf{V}^k_{i,j}.
\end{equation}
The first column of $\mathbf{V}^k_{i,j}$ is compressed at the STA then decompressed at the AP to obtain $\mathbf{\tilde{v}}^k_{i,j}$ (details of CSI compression are presented in Section~\ref{sec:csi}). For AP $j$, the CEA-ZF matrix $\mathbf{\tilde{{V}}^k_{\text{CEA}_j}} \in \mathbb{C}^{N_t \times |\mathcal{S}_j \cup \bar{\mathcal{S}}_j|}$ is constructed as

\begin{equation}
    \mathbf{\tilde{{V}}^k_{\text{CEA}_j}} = [\mathbf{\tilde{v}}^k_{1,j},\, ...,\ \mathbf{\tilde{v}}^k_{|\mathcal{S}_j|,j},\ \mathbf{\tilde{v}}^k_{1,j},\, ...,\ \mathbf{\tilde{v}}^k_{|\bar{\mathcal{S}}_j|,j}],
\end{equation}
where $\mathcal{S}_j$ is the set of scheduled in-BSS STAs of AP $j$ and $\bar{\mathcal{S}}_j$ is the set of scheduled OBSS STAs. The CEA-ZF precoder is then computed as
\begin{equation}
\mathbf{W}^k_{\text{CEA}_j} = 
(\mathbf{D}^k_{j})^{(-\frac{1}{2})}
\{\mathbf{\tilde{V}}^k_{\text{CEA}_j}((\mathbf{\tilde{V}}^k_{\text{CEA}_j})^H\mathbf{\tilde{V}}^k_{\text{CEA}_j})^{-1}\}_{[1:|\mathcal{S}_j|]},
\end{equation}
with $\{V\}_{[1:n]}$ indicating the first $n$ columns of the matrix $V$. The signal-to-interference-plus-noise ratio (SINR) at STA $i$ that is associated with AP $j$ in subcarrier $k$ is given by
\begin{equation}
\text{SINR}^k_i = 
\frac{|\mathbf{h}^k_{i,j} \mathbf{w}^k_{j,i}|^2}
{\displaystyle \sum_{s \neq i \in \mathcal{S}_j} |\mathbf{h}^k_{i,j} \mathbf{w}^k_{j,s}|^2 + \sum_{p \neq j} \sum_{r \in \mathcal{S}_p} |\mathbf{h}^k_{i,p} \mathbf{w}^k_{p,r}|^2 + N_{0} \cdot \text{NF}},
\end{equation}
Furthermore, the effective SINR is computed as the arithmetic mean across data subcarriers,
\begin{equation}
\text{SINR}^{eff}_i = \frac{1}{N_{\text{sc}}}\sum_{k=1}^{N_{sc}} \text{SINR}^k_i,
\end{equation}
and the highest MCS guaranteeing a maximum packet error rate (PER) of 10\% based on the effective SINR is selected.

\subsection{Traffic Model}
\label{sub:pkt}
Packet arrivals at the AP buffer follow a batch Poisson process (BPP). Batches arrive every interval $T_b$, modeled as $T_b \sim \text{Exp}(\lambda_b)$, with
\begin{equation}
\lambda_b = \frac{\lambda_{\text{PKT}}}{B},
\end{equation}
where $\lambda_{\text{PKT}}$ is the packet arrival rate and $B$ is the average batch size. \par

Data latency is measured as data packet delays:
\begin{equation}
    d_{\text{PKT}} = t_{\text{rec}} - t_{\text{arr}},
\end{equation}
where $t_{\text{rec}}$ and $t_{\text{arr}}$ are the reception time at the STA and the arrival time at the AP buffer, respectively. This captures queuing, contention, and transmission delays.
\section{CSI Compression and Feedback}
\label{sec:csi}

\begin{figure*}[t]
    \centering
    \includegraphics[width=\textwidth]{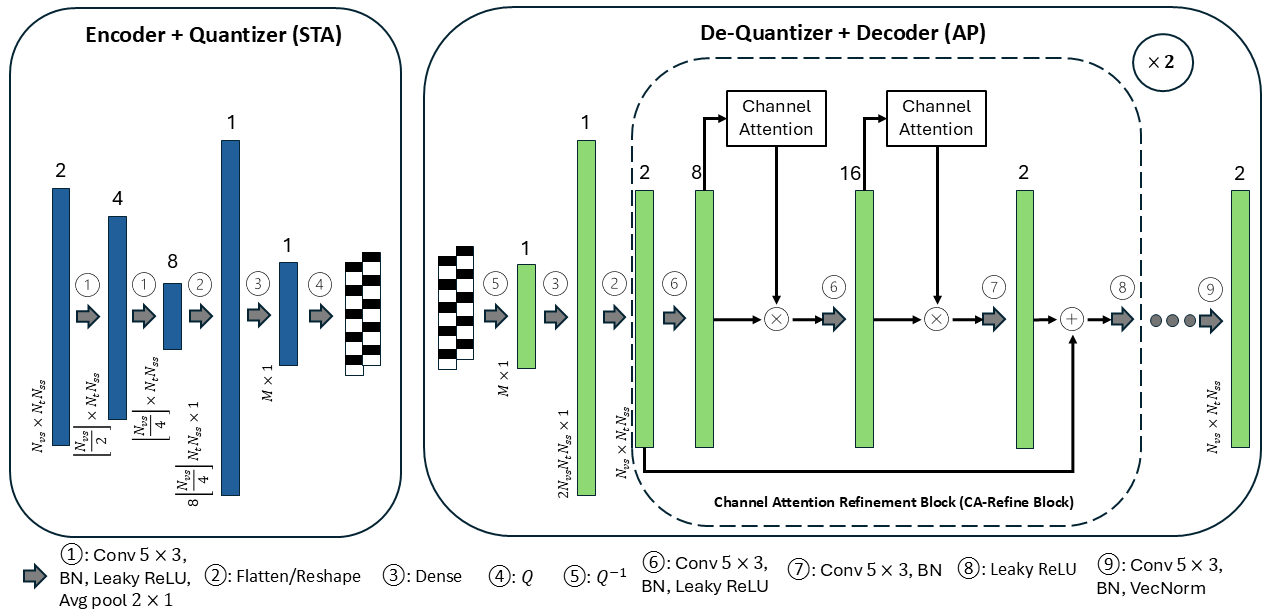}
    \caption{Proposed autoencoder architecture for CSI compression with encoder layers (blue) decoder layers (green) and entropy bottleneck layer (checkered). The number of feature maps of the input/output of each layer is indicated on top.}
    \label{fig:ae}
\end{figure*}

\begin{figure}[t]
    \centering
    \includegraphics[width=\linewidth]{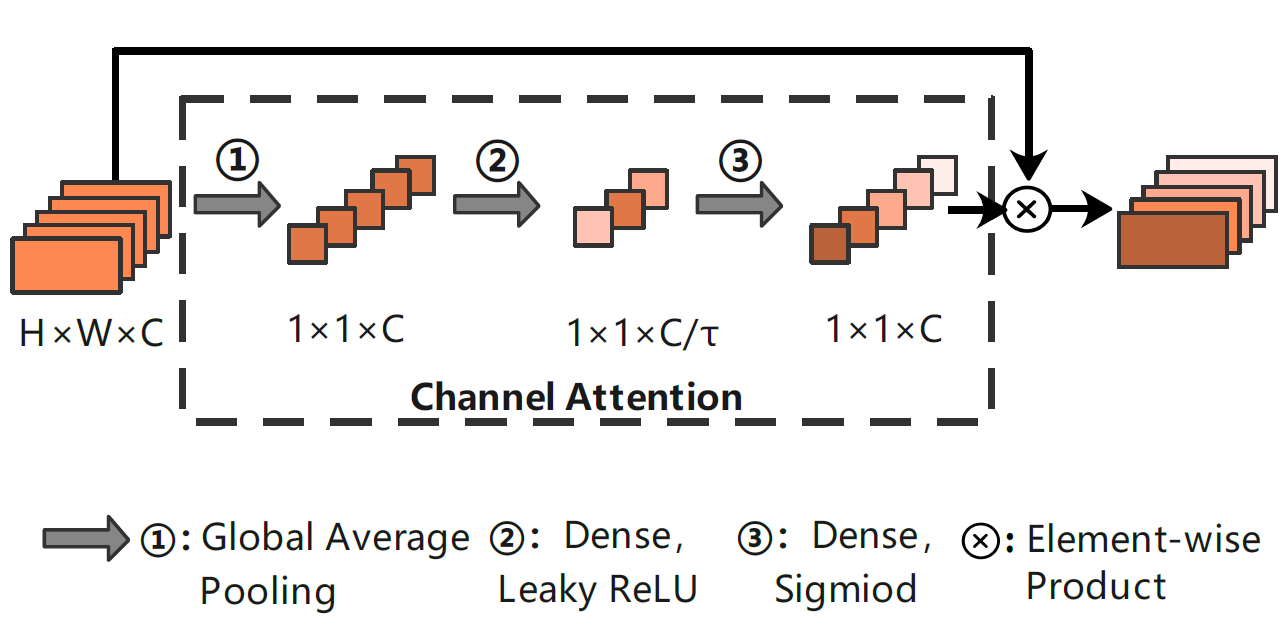}
    \caption{Architecture of the channel attention module \cite{efnet2024}.}
    \label{fig:ca}
\end{figure}

CSI compression and feedback are discussed in this section. We describe the IEEE 802.11 standard mechanism for CSI compression, followed by AE-based CSI compression. 

\subsection{CSI Compression and Feedback in IEEE 802.11}
\label{subsec:cb}

As noted in Section \ref{subsec:phy}, the right singular vectors $\mathbf{v}^k$ across multiple subcarriers are compressed at the STA and reported to the AP in the compressed beamforming (CB) report frame~\cite{ieee802.11-2024}. IEEE 802.11 uses Givens rotations to decompose each vector into angle pairs $(\phi,\psi)$, which are quantized (see Signal Model Section in~\cite{Yuen2012GR} for details). The subcarriers are grouped into blocks and the quantized angle pairs of the first subcarrier per group are reported.

The number of angle pairs is
\begin{equation}
    N_a = \frac{N_{\text{ss}}}{2}(2N_t-N_{\text{ss}}-1),
\end{equation}
where $N_{\text{ss}}$ is the number of spatial streams and $N_t$ is the number of Tx antennas. Then, the CB report size is
\begin{equation}
\left\lceil\frac{N_{\text{sc}}}{N_g}\right\rceil N_a(b_\phi+b_\psi).
\end{equation}
where $N_{\text{sc}}$ is the number of data subcarriers and $N_g$ is the subcarrier grouping. For $N_{\text{sc}}\!=\!980$, $N_t\!=\!16$, $N_g\!=\!16$, and $(b_\phi,b_\psi)\!=\!(9,7)$, the feedback size is 14,880 bits per STA, which doubles under joint
NDP sounding. This motivates AE-based compression to reduce channel sounding overheads.

\subsection{CSI Compression using Autoencoders}
\label{subsec:ae}

The proposed CSI AE is shown in Fig.~\ref{fig:ae}, which follows the EFNet design in~\cite{efnet2024} but with some enhancements on the encoder side. The unit-norm vectors $\mathbf{v}^{k}$ are sampled every $N_g$ subcarriers and stacked into an input tensor $\mathbf{T}_v \in \mathbb{R}^{2 \times N_{\text{vs}} \times N_t}$, where $N_{\text{vs}} = \lceil N_{\text{sc}}/N_g \rceil$ is the number of valid subcarriers. At the STA, $\mathbf{T}_v$ is encoded into a latent vector of dimension $M = \lceil \eta(2N_{\text{vs}} N_{t}) \rceil$, where $\eta$ is the compression ratio, which is then quantized and entropy–coded into a bit stream using an entropy bottleneck layer~\cite{Ravula2021, Balle2018}. The AP decodes and de-quantizes the latent representation, which is then expanded, reshaped and passed into two channel attention refinement (CA-Refine) blocks (see Fig.~\ref{fig:ca} for the channel attention module design) to reconstruct $\mathbf{T}_v$. A final convolution layer and per-vector normalization ensure that all reconstructed vectors have unit norm, yielding $\tilde{\mathbf{T}}_v \in \mathbb{R}^{2 \times N_{\text{vs}} \times N_t}$. Using \texttt{tanh} as in EFNet~\cite{efnet2024} bounds vector elements to $[-1,1]$ but does not ensure unit-norm vectors.

The entropy bottleneck layer follows the model of~\cite{Balle2018} and is implemented using \emph{compressai} library~\cite{begaint2020compressai}. The details of integrating the entropy bottleneck layer with AE are in Section III of~\cite{Ravula2021}. Training uses ADAM~\cite{adam2014} to minimize the mean-squared error (MSE) between $\mathbf{T}_v$ and $\tilde{\mathbf{T}}_v$ in the training set $\mathcal{D}$:
\begin{equation}
    \label{eq:mse}
    \mathcal{L}_{\text{MSE}}
    = \frac{1}{|\mathcal{D}|}\sum_{\mathbf{T}_v\in\mathcal{D}}
      |\mathbf{T}_v - \tilde{\mathbf{T}}_v|^2.
\end{equation}
To obtain $\tilde{\mathbf{v}}^{k}$ for all data subcarriers,  $\tilde{\mathbf{T}}_v$ is reshaped into unit-norm vectors, which are repeated $N_g$ times each. Reconstruction accuracy is measured using the cosine correlation between the original and reconstructed vectors:
\begin{equation}
    \rho = \frac{1}{N_{\text{sc}}} \sum_{k=1}^{N_{\text{sc}}} |(\tilde{\mathbf{v}}^k)^H \mathbf{v}^k|.
\end{equation}
\section{Performance Evaluation}
\label{sec:results}

In this section, we discuss the simulation scene, then describe the AE training data and evaluate the tradeoff between reconstruction accuracy and feedback size by examining their effect on Co-BF sounding overhead and data latency at different compression ratios. We further compare Co-BF against legacy transmissions in terms of throughput and latency.

\subsection{Simulation Scene}
\label{subsec:scene}

\begin{figure}[t]
	\centering
	\includegraphics[width=\linewidth]{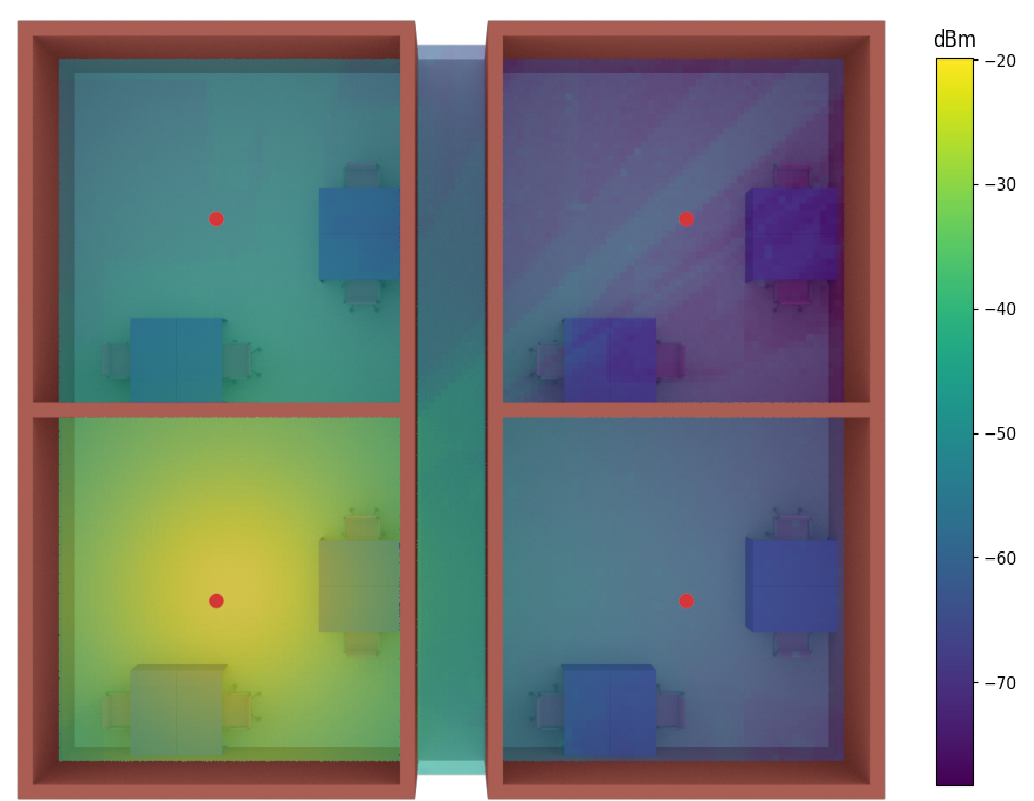}
	\caption{Indoor office with RSS heatmap of bottom-left AP.}
	\label{fig:scene}
\end{figure}

\begin{table}[t]
    \centering
    \caption{Scene and channel parameters.}
    \label{tab:phy}
    \begin{tabular}{|m{.5\columnwidth}|c|}
    \hline
    \centering\textbf{Parameter}  & \textbf{Value} \\
    \hline
    Room length, width, height [m]  & (5, 5, 3) \\
    Corridor width [m] & 1 \\
    STA height [m] & 1.2 - 1.7 \\
    STA speed [km/h] & 0.9 \\
    Channel bandwidth [MHz] & 80 \\
    Subcarrier spacing [kHz] & 78.125 \\
    Carrier frequency [GHz] & 5.3 \\
    Noise floor [dBm/Hz] & -174 \\
    Power spectral density [dBm/MHz] & 5 \\
    Noise figure (NF) [dB] & 7 \\
    \hline
    \end{tabular}
\end{table}

The scene used for AE training and event-driven simulations is presented in Fig.~\ref{fig:scene}, which is an indoor office scene with 4 rooms. Each room in the scene contains a ceiling-mounted AP with a 4$\times$4 uniform planar array (UPA) antenna with a directional radiation pattern, while STAs use single omnidirectional antennas. A received signal strength (RSS) heatmap of one of the APs over the scene is also shown in Fig.~\ref{fig:scene}. The STAs are deployed at random heights with at least 50 cm horizontal separation from the walls. Furthermore, during event-driven simulations, the STAs move inside the scene at a fixed speed in random directions (random-walk), with the direction of movement changing every 2 seconds. The scene and channel parameters are listed in Table~\ref{tab:phy}. \par

For event-driven simulations, we consider multiple deployment scenarios with different transmission modes, CSI compression methods, and traffic loads. For each scenario, the STAs are placed in two randomly selected rooms of the scene in Fig.~\ref{fig:scene}, each served by the AP in that room. Per-STA traffic loads are 177/93/63 Mbps for 2/4/6 STAs per AP (high load),\footnote{High loads were set at 95\% of worst-case STA throughput of legacy 80.} with medium load set to half these values. Each setting is evaluated in 50 deployments using 10-second simulations with identical seeds for topology, mobility, and traffic generation to ensure a fair comparison between different settings. \par

\subsection{Co-BF Evaluation with AE CSI Compression}
\label{subsec:csi_delay}

\begin{table}[t]
\caption{Reconstruction accuracy and feedback size.}
\label{tab:corr}
\centering
\scriptsize
\renewcommand{\arraystretch}{1.1}
\setlength{\tabcolsep}{2pt}
\begin{tabular}{l cc cc cccc}
\hline
\multirow{2}{*}{\textbf{Dataset}}
  & \multicolumn{2}{c}{\textbf{Mean}}
  & \multicolumn{2}{c}{\textbf{1st Percentile}}
  & \multicolumn{4}{c}{\textbf{Feedback size [bits]}} \\
 & \textbf{LOS} & \textbf{NLOS}
 & \textbf{LOS} & \textbf{NLOS}
 & \textbf{Median} & \textbf{Min} & \textbf{Max} & \textbf{StDev} \\
\hline
IEEE 802.11 (conf. 1) & \textbf{0.9998} & \textbf{0.9861} & \textbf{0.9994} & 0.8609 & 14880 & -- & -- & -- \\
IEEE 802.11 (conf. 2) & 0.9988 & 0.9880 & 0.9986 & \textbf{0.8886} & 44100 & -- & -- & -- \\
\hline
Train ($\eta=1/2$) & 0.9975 & 0.9756 & 0.9951 & 0.7857
 & \multirow{2}{*}{6240} & \multirow{2}{*}{5760} & \multirow{2}{*}{7072} & \multirow{2}{*}{205.19} \\
Test ($\eta=1/2$)  & 0.9974 & 0.9748 & 0.9938 & 0.7846
 & \multicolumn{4}{c}{} \\
\hline
Train ($\eta=1/4$) & 0.9974 & 0.9758 & 0.9931 & 0.7715
 & \multirow{2}{*}{2944} & \multirow{2}{*}{2784} & \multirow{2}{*}{3232} & \multirow{2}{*}{75.20} \\
Test ($\eta=1/4$)  & 0.9970 & 0.9745 & 0.9885 & 0.7957
 & \multicolumn{4}{c}{} \\
\hline
Train ($\eta=1/8$) & 0.9931 & 0.9718 & 0.9783 & 0.7887
 & \multirow{2}{*}{1472} & \multirow{2}{*}{1408} & \multirow{2}{*}{1600} & \multirow{2}{*}{28.78} \\
Test ($\eta=1/8$)  & 0.9914 & 0.9634 & 0.9598 & 0.7713
 & \multicolumn{4}{c}{} \\
\hline
Train ($\eta=1/16$) & 0.9853 & 0.9662 & 0.9505 & 0.7859
 & \multirow{2}{*}{\textbf{768}} & \multirow{2}{*}{\textbf{736}} & \multirow{2}{*}{\textbf{864}} & \multirow{2}{*}{21.36} \\
Test ($\eta=1/16$)  & 0.9821 & 0.9564 & 0.9077 & 0.7480
 & \multicolumn{4}{c}{} \\
\hline
\end{tabular}
\end{table}

\begin{figure}[t]
    \centering
    \includegraphics[width=\linewidth]{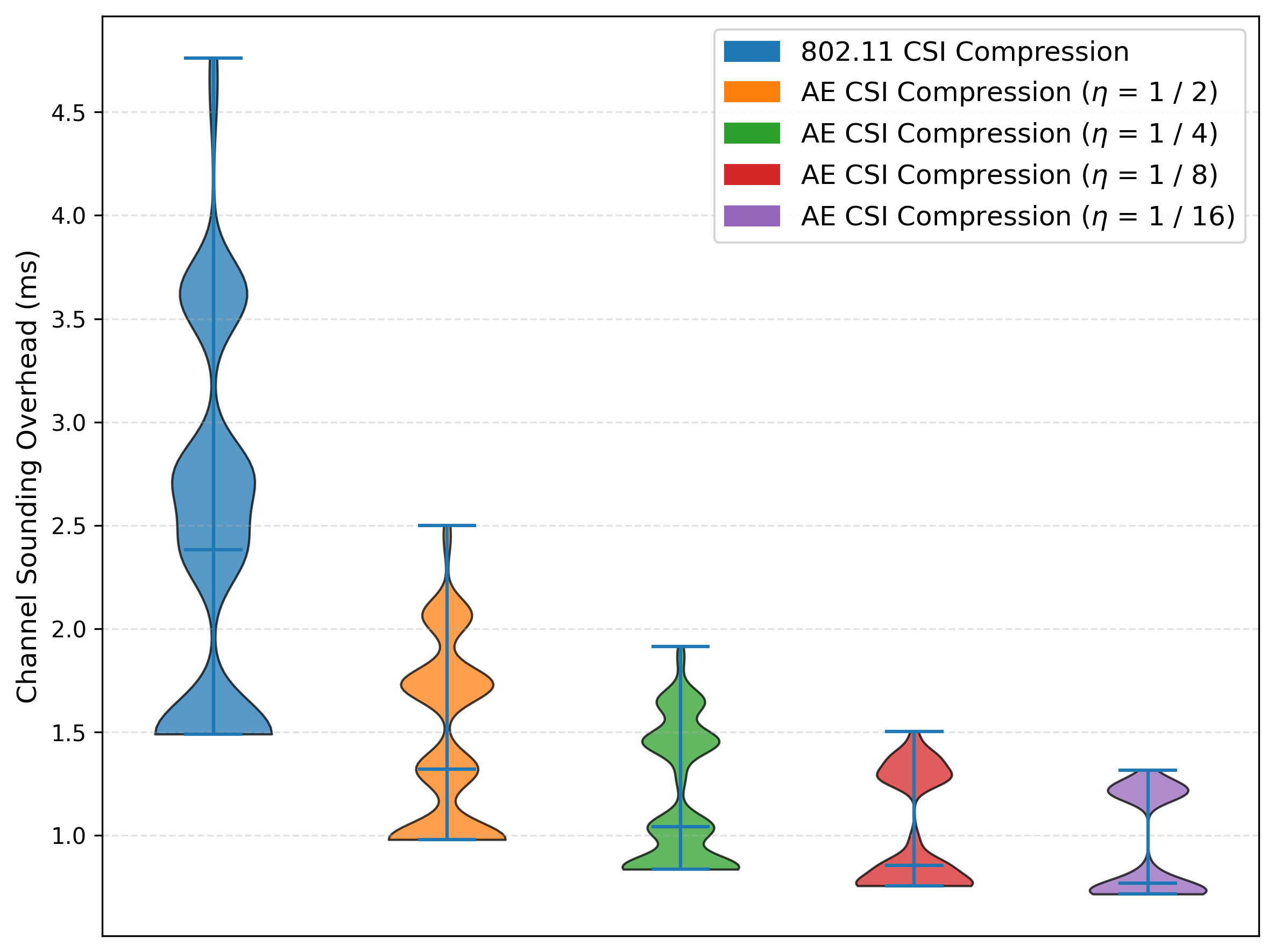}
    \caption{Violin plots of Co-BF channel sounding overhead at high load.}
    \label{fig:csi}
\end{figure}

\begin{figure}[t]
    \centering
    \includegraphics[width=\linewidth]{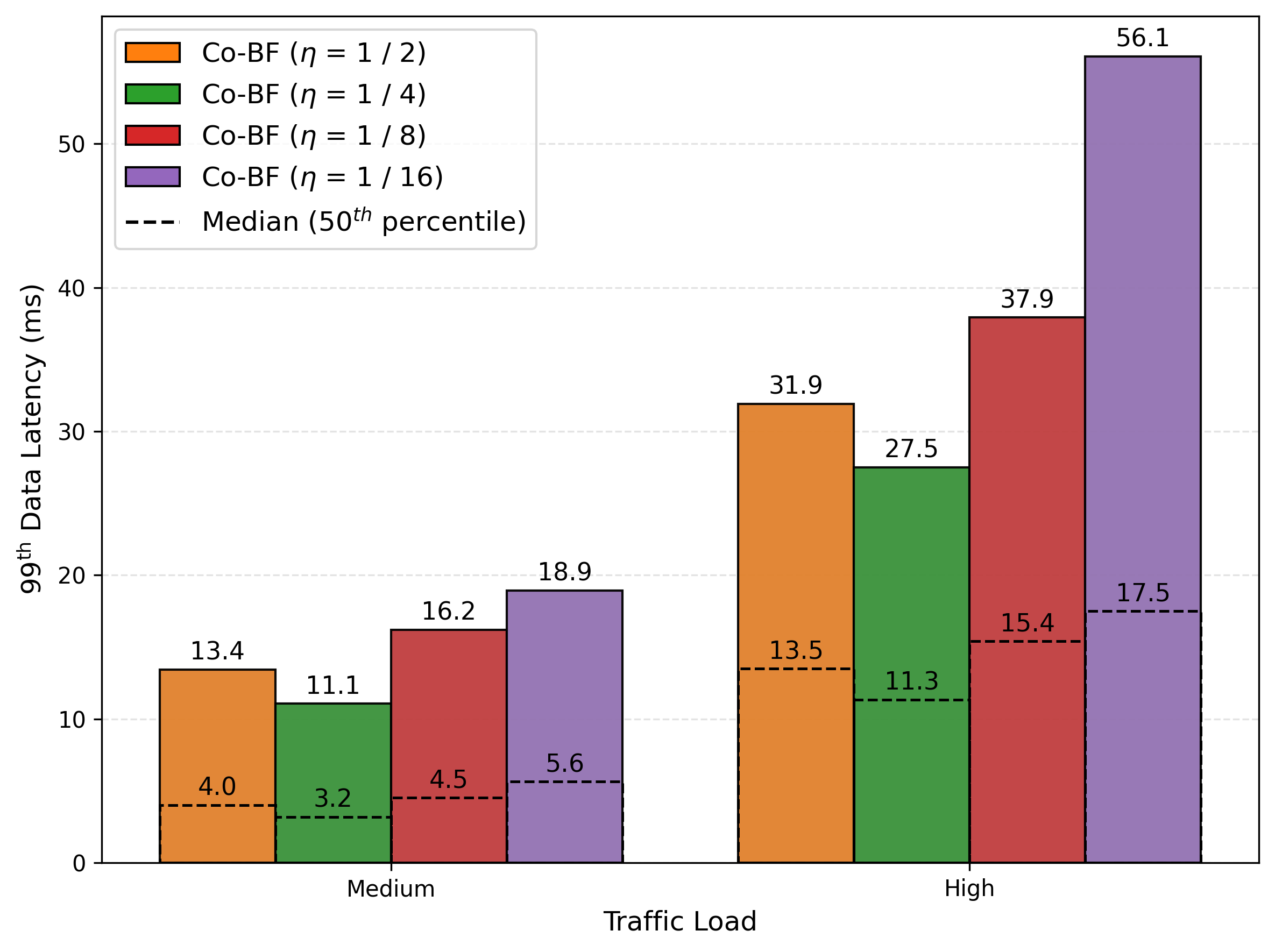}
    \caption{$99^{th}$ percentile and median latency with 4 STAs per AP.}
    \label{fig:pkt_ae}
\end{figure}

\subsubsection{AE Training Data}
\label{subsub:ae_data}

The AE is trained using CSI collected from the scene in Fig.~\ref{fig:scene} via Sionna RT. We collected 12,000 CSI samples per room (four APs at 3000 random STA positions). For Table~\ref{tab:corr}, the training uses three rooms (36,000 samples), while the remaining room provides validation and test data to assess generalization between rooms. For the performance evaluations in this section, we use a final AE trained on the entire dataset.

\subsubsection{AE Evaluation}
Table~\ref{tab:corr} reports the mean and bottom 1st percentile reconstruction accuracy under LOS and NLOS conditions and the feedback sizes for various $\eta$. Furthermore, two IEEE 802.11 CSI feedback configurations are included for reference: configuration 1 with the baseline parameters from Section~\ref{subsec:cb}, and configuration 2 with $N_g=4$ and $(b_{\phi}, b_{\psi})=(7,5)$. \par

Reconstruction accuracy is similar for $\eta = 1/2$ and $\eta = 1/4$, but drops at lower ratios, while feedback size decreases almost linearly with $\eta$. In addition, the AE shows very good generalization between different rooms, which is evident from the almost identical accuracy between the training and testing sets. The generalization is best with $\eta = 1/2$ and $\eta = 1/4$, but decreases with lower $\eta$. Nevertheless, the 4 rooms in the scene have very similar channel distributions, and further generalization capabilities between different scenarios shall be studied in future work. Furthermore, Figures~\ref{fig:csi} and~\ref{fig:pkt_ae} illustrate the system-level impact. Fig.~\ref{fig:csi} shows that IEEE 802.11 compression can push the sounding overhead near the 5.484 ms TXOP limit, while AE compression keeps it well below, allowing sufficient time for A-MPDU transmission after channel sounding. Fig.~\ref{fig:pkt_ae} indicates that $\eta = 1/4$ offers the lowest data latency by balancing reconstruction accuracy and feedback size: it halves the feedback relative to $\eta = 1/2$ at nearly the same reconstruction accuracy and avoids the degraded precoder–channel alignment seen at $\eta < 1/4$.

\subsection{Co-BF Comparison with Legacy}
\label{subsec:pkt_delay}

\begin{figure}
    \centering
    \includegraphics[width=\linewidth]{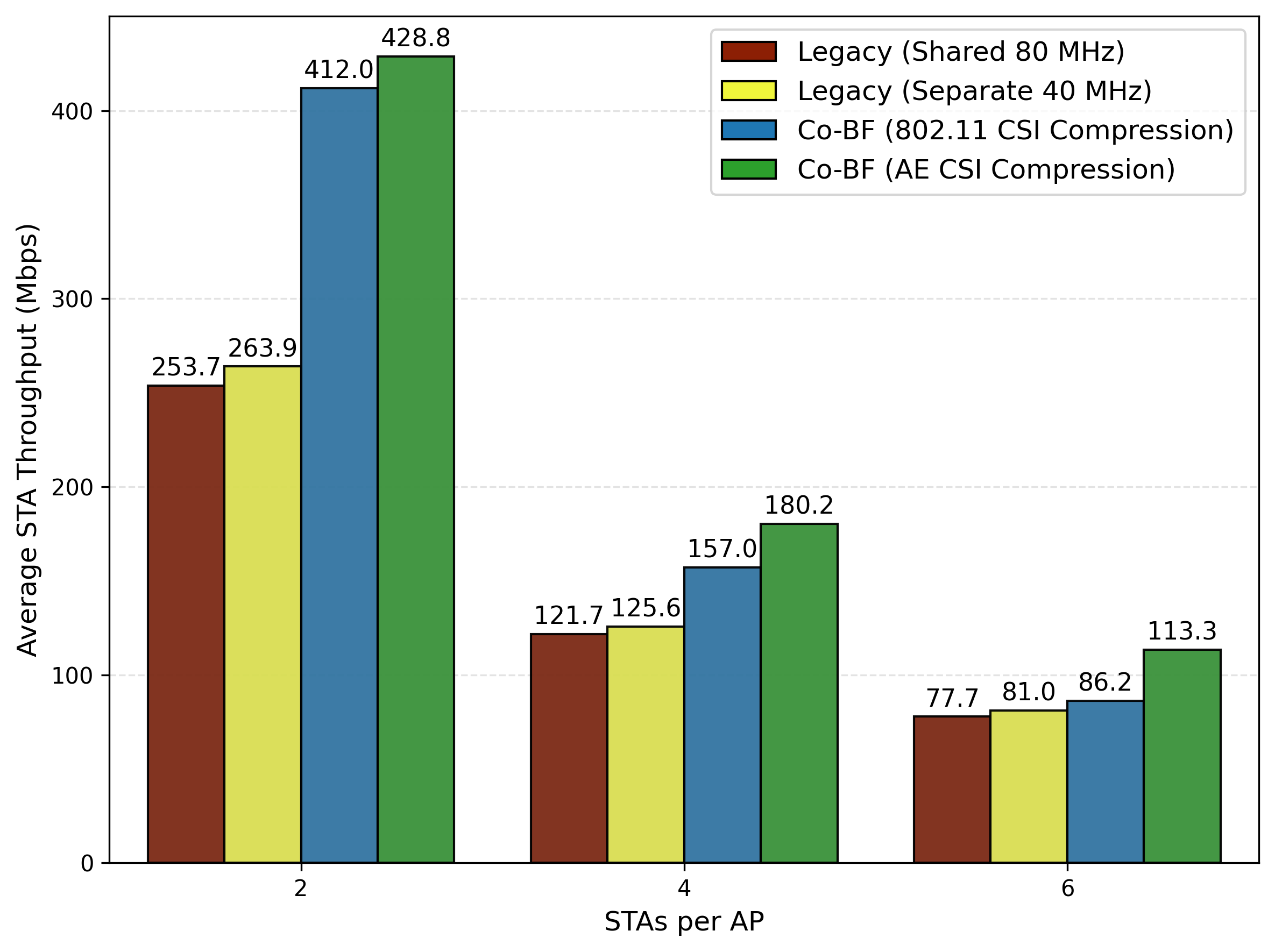}
    \caption{Average throughput per STA comparison.}
    \label{fig:throughput}
\end{figure}

\begin{figure}[t]
    \centering
    \includegraphics[width=\linewidth]{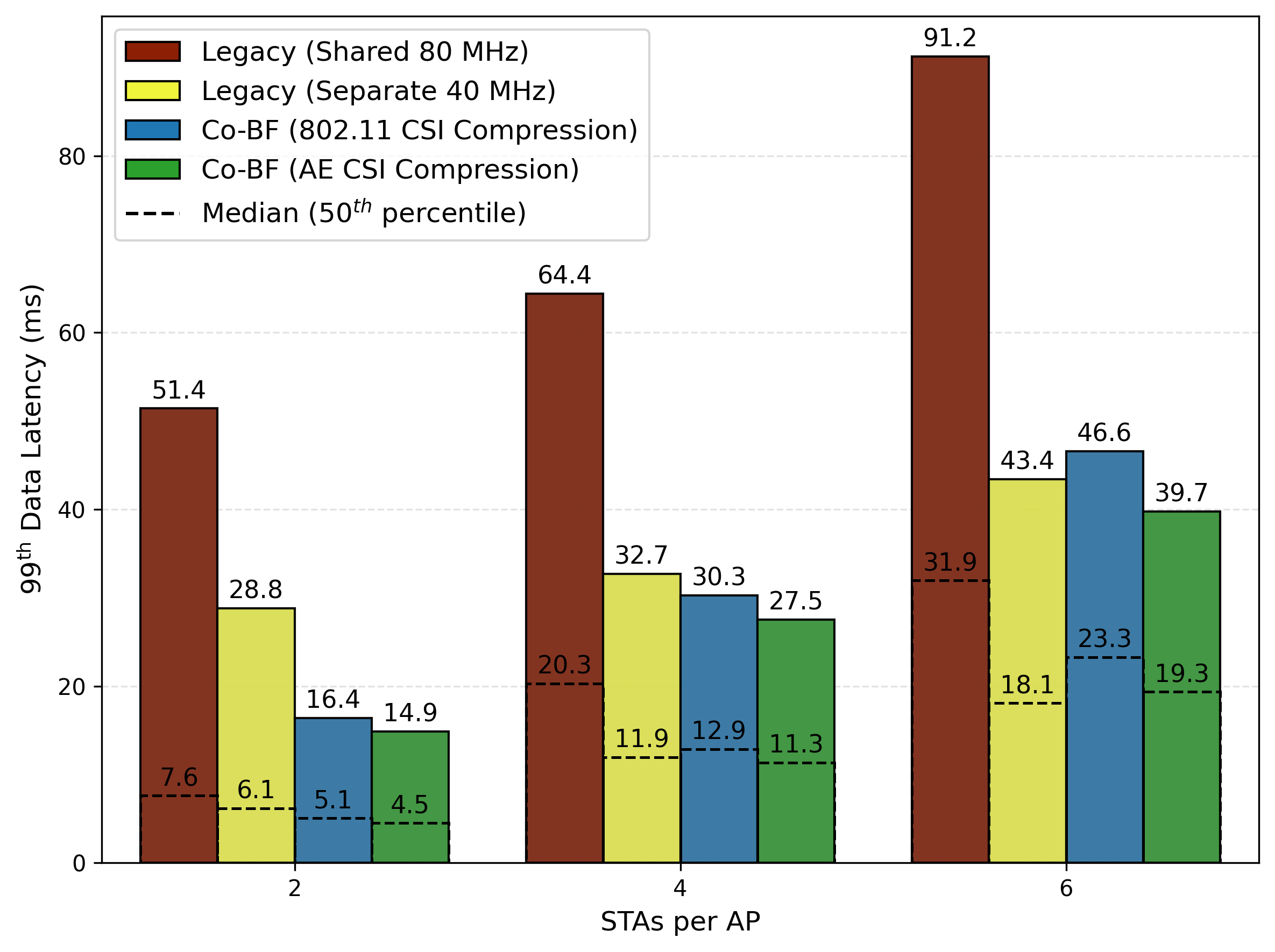}
    \caption{$99^{th}$ percentile and median data latency comparison.}
    \label{fig:pkt_all}
\end{figure}

Figures~\ref{fig:throughput} and~\ref{fig:pkt_all} show comparisons of Co-BF and legacy non-MAPC performance in terms of throughput (full buffer) and data latency (high load), respectively. Co-BF is evaluated with IEEE 802.11 CSI compression and AE-based CSI compression, which we will refer to as $\text{Co-BF}_{\text{ST}}$ and $\text{Co-BF}_{\text{AE}}$, respectively. In addition, legacy transmissions are evaluated with both APs using a shared 80~MHz channel or each using a separate 40~MHz channel, which we will call legacy 80 and legacy 40, respectively. With 2 STAs per AP, Co-BF increases throughput by more than 50\% compared to the legacy 40 MHz case using either CSI compression technique. There are marginal gains in throughput and data latency with $\text{Co-BF}_{\text{AE}}$ compared to $\text{Co-BF}_{\text{ST}}$, but the performance is already well improved with Co-BF at this STA density.

With 4 STAs per AP, Co-BF offers marginal latency reductions but with decent throughput gains compared to the legacy, which are further increased by AE CSI compression. However, $\text{Co-BF}_{\text{ST}}$ results in higher latency than legacy 40 with 6 STAs per AP, which is caused by large sounding overheads. However, $\text{Co-BF}_{\text{AE}}$  reduces the $99^{th}$ latency compared to legacy 40 and also produces a significant increase of 39.9\% in throughput. The median latency is still slightly higher with $\text{Co-BF}_{\text{AE}}$ compared to legacy 40, but this is due to the absence of collisions and minimal contention overhead in legacy 40. These results show that AE CSI compression is an effective tool for reducing sounding overhead, significantly improving Co-BF performance across a wide range of STA densities and traffic conditions.
\section{Future Work}
\label{sec:future}

The results presented show that using AE for CSI compression can significantly boost the performance of Co-BF. However, the AE in this work is trained with site-specific data from the same simulation scene. Generalization between different scenes has been studied for cellular systems~\cite{Song2024, Lou2025}, but the evaluation is limited to the reconstruction accuracy of the channel. The impact of generalization on E2E performance has not yet been studied. In addition, different considerations for model architectures may be required for Wi-Fi channels, since most current AE architectures were built for 3GPP channels, not Wi-Fi channels. \par

AE generalization for different channel models, bandwidths, and antenna sizes in Wi-Fi was studied in IEEE contributions~\cite{Ziyang_Gou2023}, but they lack results of the impact on E2E performance. Wi-Fi-based studies on AE CSI compression are also limited for SU-MIMO. A universal AE model that generalizes to different channel distributions, bandwidths, and antenna configurations must be developed and tested in MU-MIMO/Co-BF scenarios for practical deployment. Model size, sharing, and run-time are other crucial aspects that have to be studied for practical use of AEs for CSI compression.
\section{Conclusion}
\label{sec:conclusion}

We presented an AE-based CSI compression mechanism to reduce the large channel sounding overhead of Co-BF for next-generation Wi-Fi. Using a standards-aligned MAC design and realistic Sionna RT channels, we showed that AE compression substantially reduces channel sounding overhead by over 50\%, and provides an optimal accuracy/feedback-size tradeoff at $\eta = 1/4$ in terms of data latency performance.

We further evaluated Co-BF against non-MAPC legacy transmissions and saw that IEEE 802.11 CSI compression limits Co-BF performance, leading to higher latency than the legacy at high STA densities. However, Co-BF with AE-based CSI compression outperforms the legacy in all cases with significant gains in throughput and latency. These results indicate that AE-based CSI compression is a strong candidate for enabling practical and high-performance MAPC operation in Wi-Fi 8 and beyond.

\bibliographystyle{IEEEtran}
\bibliography{references}

\end{document}